# Performance Improvement of VOD System by Policy Based Traffic Handle

Soumen kanrar

*Abstract*—The distributed system use to enhance the performance of all types of multimedia service in the next generation network. The packet loss occurs in the video on demand system due to delay and huge traffic load from the both sides of client request and server response. It's bring a real challenge to the researcher how to minimize the traffic load in the video on demand system to provide better utilization of the available bandwidth in the very race situation. The normal client server type architecture can't solve the huge client requirement. In this work, I have used number of remote servers that partly shears the total load in the video on demand system on behalf of the total system load. This work presents the scenario of controlling the traffic load in the localization wise sub domain that gives good impact to control the traffic in the whole distributed system. The user increased rapidly in the network it posed heavy load to the video servers. The requested clients, servers are all distributed in nature and the data stream delivered according to client request. In this work presents the performance of the video on demand server by policy based traffic handle, at real time with respect to incoming multirate traffic pattern.

## I. INTRODUCTION

A typical VoD architecture consists of three critical subsystems: single or cluster video servers, high speed wide-area and local distributed networks and many users' populations. The hierarchical VoD system architecture used in this analysis consist of local and remote sites. Each site is characterized by a cluster of video servers. The video servers deliver high quality digitized multimedia data to clients over local distributed networks from local sites or over high speed networks from remote sites. The set-top boxes at the client site provide the decoding and display functionality, in addition to providing buffer for periodically delivered video segments from the video servers.

Reference number, as in [1] used to the video on demand system to get better performance. Planning is very important reference number, as in [2], [3] to properly manage the resources. The video on demand system are of any type's networks, .i.e. confined or isolated or distributed. The network faced huge traffic loss and packet delayed due to the bottleneck problem reference number, as in [4]-[6]. The client request is exponentially increases with respect to time.

It is very important to provide similar or better performance to the client with the consistent packet traffic flow in the network. Authors, reference number, as in [7], [8], [9] proposed to used the cluster concepts for handling the packet loss and congestion level. The distributed network provides better performance with respect to the latency time and packet loss. So and width optimization regarding these types of problem is further to cultivate. For the heterogeneous types of network channel allocation to the user is an issue for the limited bandwidth reference number, as in [10]. In order to support a large population of clients, the network need new types of better solutions that efficiently utilize the server and network resources.

The request comes from the client to the VoD server for the different types of request. Main categories are classified in two broad areas. One for the popular clips and other for the unpopular clips. The request is also for two types in the VoD Network; the first one of the request for initializing or starting the video clips. The other type is the request for interactive service (e.g. stop/pause, jump forward, fast reverse etc) to be performed on the viewed clips. Since each of these request is independent from each other, and arrival requests come from large numbers of client set-up terminals, the arrival process of normal requests as well as of interactive requests to the video server can be modeled as a Poisson distribution with average rates $\lambda_S$ ( for steady session) and $\lambda_I$ ( for interactive session) respectively. With this assumption, the distribution of the sum of $K$ of independent identically distributed random variables, representing the request inter – arrival times (Which are exponential distributed mutually independent random variables) is then follow the Erlang distribution.

Multicast can significantly improve the VoD system performance greatly by reducing the required network bandwidth. So the overall network load reduces. In other way the multicasting alleviates the workload of the VoD server and improves the system throughput by batching requests. Multicasting offer excellent scalability which in turn, enables serving a large number of clients that provide excellent cost/performance benefits. In this paper I have shown how the traffic load to the server can be control by using the policy based Traffic handle. I have presented the significant impact of policy based Traffic handle on the performance of the VoD system. This paper is structured as follows. Section 1 introduces the main text and the introduction of video on demand in the distributed system and related issues. The next section 1.1 presents the analytic model. Section 1.2 briefly focused on the simulation parameters and section 1.3 shows the obtained results and followed by corresponding discussion with respect to the proposed policy based Traffic handle.

## II. ANALYTIC MODEL

Let the request comes from $i^{th}$ class of population and served by the $j^{th}$ partition block of the server where $1 < j$ and assume $A_j B_i B_{i+1} \ldots \ldots B_{j-1}$ is the event that the previous all partition from $i$ to $j-1$ is blocked only $j^{th}$ partition has at least one free port.

$B_i B_{i+1} \ldots \ldots B_{j-1}$ represents the event that all the partition from $i$ to $j-1$ is blocked.

We get,

$$p(A_j / B_i B_{i+1} \ldots \ldots B_{j-1})$$
$$= \frac{p(A_j B_i B_{i+1} \ldots \ldots B_{j-1})}{p(B_i B_{i+1} \ldots \ldots B_{j-1})}$$
$$= \frac{p(A_j) p(B_i) p(B_{i+1}) \ldots \ldots p(B_{j-1})}{p(B_i) \ldots \ldots p(B_{j-1})}$$
$$= p(A_j)$$
$$= (1 - \frac{1}{k})^{j-1} \cdot \frac{1}{k} (\frac{C_j - Q_i^{(j)}}{C_j}) \quad (1)$$

Since, $p_b(B_m \cap B_n) = \Phi$, $\forall (m,n) \in N$ and $m \neq n$ where $i \leq m \leq j-1, i \leq n \leq j-1$.

So we get

$p_b(B_i B_{i+1} \ldots \ldots B_{j-1}) = p_b(B_i) p_b(B_{i+1}) \ldots \ldots p_b(B_{j-1})$
$\ldots \ldots \ldots (11)$

Each $p_b(B_i)$ for $i \in I^{\geq o}$ follows the Erlang distribution.

So the above expression (11) can be represent as
$p_b(B_i B_{i+1} \ldots \ldots B_{j-1}) =$

$$\left[ \frac{\frac{E_i^{C_i}}{(c_i)!}}{\sum_{k=0}^{c_i} \frac{E_i^k}{(k)!}} \right] \cdot \left[ \frac{\frac{E_{i+1}^{C_{i+1}}}{(c_{i+1})!}}{\sum_{k=0}^{c_{i+1}} \frac{E_{i+1}^k}{(k)!}} \right] \ldots \ldots \ldots \left[ \frac{\frac{E_{j-1}^{C_{j-1}}}{(c_{j-1})!}}{\sum_{k=0}^{c_{j-1}} \frac{E_{j-1}^k}{(k)!}} \right] \quad (2)$$

Now the request from the class a will be admitted to a partition b with probability

$$p^{\#}(a,b) = p_a^{\#} \quad (3)$$

where $\sum_{i=1}^{k} p^{\#}_i = 1$, A new request arrives with Poisson distribution for video stream.

$B_i B_{i+1} \ldots \ldots B_{j-1}$ represents the event that all the partition from $i$ to $j-1$ is blocked. Now by considering the expression (1) we get,

$$p(A_j / B_i B_{i+1} \ldots \ldots B_{j-1})$$
$$= (p_a^{\#}) \frac{p(A_j B_i B_{i+1} \ldots \ldots B_{j-1})}{p(B_i B_{i+1} \ldots \ldots B_{j-1})}$$
$$= (p_a^{\#}) \frac{p(A_j) p(B_i) p(B_{i+1}) \ldots \ldots p(B_{j-1})}{p(B_i) \ldots \ldots p(B_{j-1})}$$
$$= (p_a^{\#}) p(A_j) \quad (4)$$

## III. EXPLANATION OF SIMULATION ENVIRONMENT

The simulation environment is created according to the problem requirement.

For the simplicity, I have considered that each sector of the server contained equal number of ports. The traffic arrived rate from different cluster of clients, started at 1 Mb/s and end at 5 Mb/s. The number of clusters of client is 30. The number of subsection in the server side is 30. Port access time for each client vary from 1sec to 30 second for each video clips. The simulation runs for 500 seconds. Table 1 presents the overall simulation parameters for the VoD server.

TABLE I: SIMULATION PARAMETERS

| The Simulation Parameters of VoD Server | Parameters - Value |
|---|---|
| Number of Clusters | 30 |
| Minimum traffic Arrival rate | 1 Mb/sec |
| Maximum Traffic Arrival Rate | 15.5 Mb/sec |
| Port access time minimum | 1 sec |
| Port access time maximum | 200 sec |
| Number Sub section | 30 |
| Simulation Time | 500 seconds |

## IV. PERFORMANCE ANALYSIS

Figure 1 to 5 represents the performance of distributed video on demand system with respect to the policy based traffic handle. Figure.1 (a) and (b) represents the video on demand system without any traffic handle. The incoming traffic flows from the 30 clusters of clients. The rate of the traffic started 1Mb/second from the first cluster. 1.5 Mb/second from the second cluster and 15.5 Mb/second from the 30$^{th}$ client cluster.

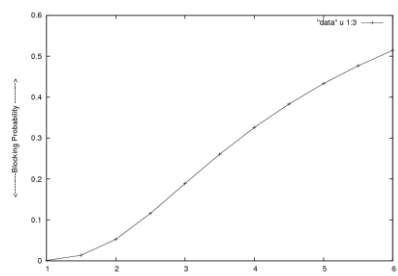

Fig. 1. (a) Traffic arrival / Blocking probability ;

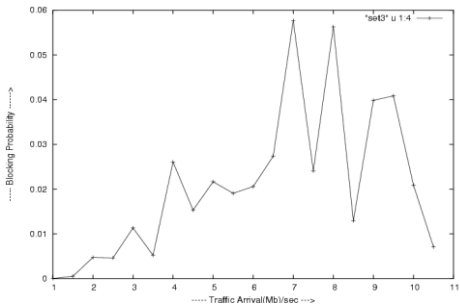
Fig. 2. (c) Traffic arrival /Blocking probability.

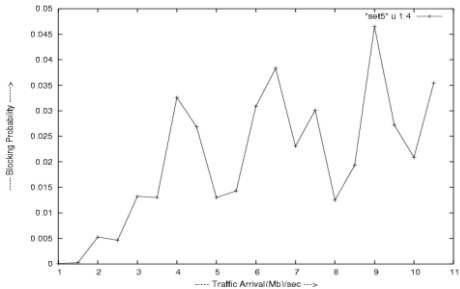
Fig. 3. (e) Traffic arrival /Blocking probability.

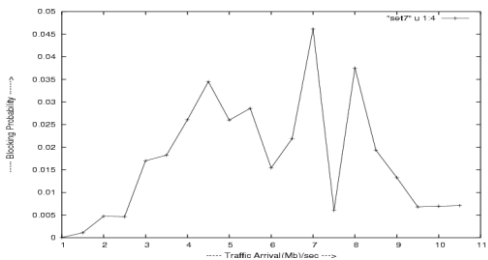
Fig. 4. (g) Traffic arrival /Blocking probability.

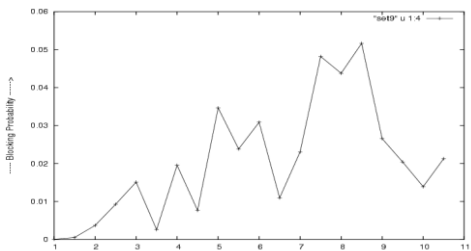
Fig. 5. (i) Traffic arrival /Blocking probability.

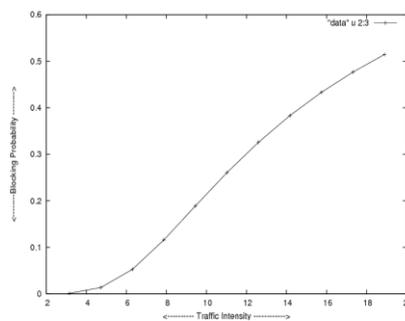
Fig. 1. (b) Traffic intensity / Blocking probability .

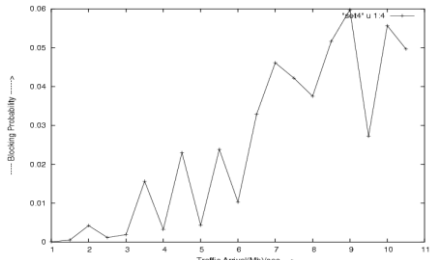
Fig. 2. (d) Traffic arrival /Blocking probability.

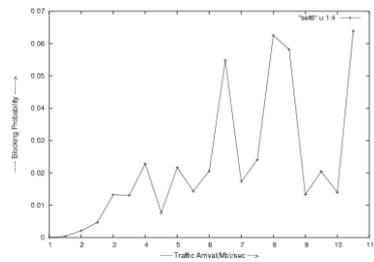
Fig. 3. (f) Traffic arrival /Blocking probability

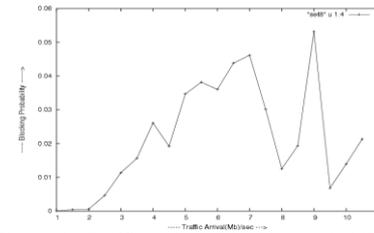
Fig. 4. (h) Traffic arrival /Blocking probability

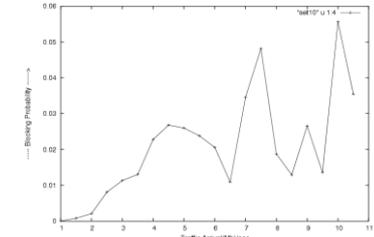
Fig. 5. (j) Traffic arrival /Blocking probability

The figure 2(c) to figure 5(j) represents the performance of the distributed video on demand system with respect to the control probability set, for the incoming traffic rate in the distributed video on demand system. The simulation result represents for any time or in any high demand situation, the blocking probability is below some threshold level. If we compare for any incoming traffic rate via the figure 1(a) and (b) with the figures 2(c) to 5(j). The policy based traffic handle well control the traffic congestion in the distributed network as well as it enhanced the overall system performance.

*International Journal of Modeling and Optimization, Vol. 2, No. 3, June 2012*

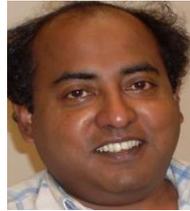

**S. Kanrar** received the M.Tech. degree in computer science from Indian Institute of Technology Kharagpur India in 2000. Advanced Computer Programming RCC Calcutta India 1998. and MS degree in Applied Mathematics from Jadavpur University India in 1996. BS degree from Calcutta University India. Currently he is working as researcher at Vehere Interactive Calcutta India. Formally attached with the University Technology Malaysia. He is the member of IEEE. Email: Soumen.kanrar@veheretech.com